\documentclass[11pt]{article}

\usepackage[margin=2cm]{geometry}
\usepackage{bm}
\usepackage{overpic}
\usepackage{amsmath}
\usepackage{amssymb}
 \usepackage{doi}

\begin{document}

\title{Drop spreading with random viscosity}

\author{Feng Xu and Oliver E. Jensen \\ 
School of Mathematics, University of Manchester, \\ Oxford Road, Manchester M13 9PL, UK}

\maketitle

\begin{abstract}
We examine theoretically the spreading of a viscous liquid drop over a thin film of uniform thickness, assuming the liquid's viscosity is regulated by the concentration of a solute that is carried passively by the spreading flow.   The solute is assumed to be initially heterogeneous, having a spatial distribution with prescribed statistical features.  To examine how this variability influences the drop's motion, we investigate spreading in a planar geometry using lubrication theory, combining numerical simulations with asymptotic analysis.  We assume diffusion is sufficient to suppress solute concentration gradients across but not along the film.  The solute field beneath the bulk of the drop is stretched by the spreading flow, such that the initial solute concentration immediately behind the drop's effective contact lines has a long-lived influence on the spreading rate.  Over long periods, solute swept up from the precursor film accumulates in a short region behind the contact line, allowing patches of elevated viscosity within the precursor film to hinder spreading.   A low-order model provides explicit predictions of the variances in spreading rate and drop location, which are validated against simulations.
\end{abstract}

\section{Introduction}

The thin liquid film lining lung airways plays an important role in protecting airway tissues from the harmful effects of inhaled particles or aerosol droplets \cite{thornton2008structure, fahy2010airway, levy2014pulmonary}.  The film is a complex liquid that includes mucins, surfactants and surfactant-associated proteins; its thickness is regulated by osmotic effects driven by ion fluxes across airway epithelial cells and its transport is driven by active motion of cilia on epithelial cells.  The film's rheology is dependent in part on the secretion of mucins from goblet cells distributed across the airway wall; disruption of normal mucin production can lead to harmful effects associated with poor clearance of pathogens.  The physical properties of the film in a particular airway of a given individual are therefore subject to considerable uncertainties and intrinsic spatial variability \cite{lai2009micro, didier2012statistical}.
 
These features motivate the present study, in which we seek to relate the spatial heterogeneity of a liquid film to the dynamics of a drop spreading over it.  We deliberately focus on a subset of features relevant to airway liquid, neglecting non-Newtonian rheology, osmotic effects, internal stratification and ciliary transport.  Instead we assume that the film's viscosity is determined by the concentration of a solute (a proxy for mucins, strong determinants of mucus viscosity \cite{georgiades2014particle}) that is distributed heterogeneously and diffuses slowly within the film.  We wish to establish how spatial variability in solute concentration influences the rate at which an inhaled aerosol droplet might spread over the film.  This allows us to address an equivalent, related question: given imperfect knowledge of the film's properties, what is the likely distribution of spreading rates?
 
To investigate these questions, we exploit a sequence of approximations.  The drop and the film over which it spreads are both assumed to be thin, allowing the flow to be described using lubrication theory.   We assume the solute is of an appropriate molecular weight to diffuse across the film during the lifetime of the drop spreading, but not appreciably along it.  This allows us to simulate the spreading flow using a pair of coupled transport equations, for the film thickness and the cross-sectionally averaged solute distribution.  The initial solute distribution along the film is described as a Gaussian random field with a specified covariance.  We can then simulate multiple (Monte Carlo) realisations of spreading dynamics, although this is computationally expensive.  Further progress can be made by assuming the drop height significantly exceeds the precursor film thickness.  As is well known from numerous studies (reviewed in \cite{bonn2009wetting, snoeijer2013moving, sibley2015comparison, eggers2015singularities}), the drop dynamics is then regulated by the flow in narrow `inner' regions in the neighbourhood of the drop's effective contact lines.  Analysis using characteristics shows that the solute distribution ahead of the drop is swept into each inner region where it concertinas as the drop advances over the film; in contrast, the solute distribution within the remainder of the drop is stretched by the spreading.  At any instant, the drop spreading rate is regulated primarily by the viscosity at the rear of each inner region, where it overlaps with the bulk `outer' region.  We exploit these observations to derive a set of nonlinear ODEs (a surrogate of the full system) that captures drop spreading rates and which allows the statistical variability in the dynamics to be characterised efficiently.  Further simplifications arise when the disorder in the initial viscosity field is weak.

Our study complements numerous previous theoretical studies of drop spreading and contact-line motion.  The precursor film regularises the contact-line singularity \cite{chebbi1999, kalinin1986}; we avoid introducing slip or a disjoining pressure, while recognising that these may be relevant in some applications.  While there are numerous potential origins of randomness (thermal fluctuations \cite{grun2006thin}, a rough surface \cite{miksis1994slip, cox1983spreading, krechetnikov2005experimental, bonn2009wetting, savva2009, savva2010two}, etc.), we focus here on spatial heterogeneity of the liquid itself, a feature that is particularly relevant to biological applications.  The problem is governed by four primary dimensionless quantities (a precursor film thickness; a P\'eclet number; the variance and correlation length of the initial random field); rather than attempt comprehensive coverage of parameter space, we investigate distinguished limits in which insights are possible through model reduction techniques.  

\section{The model problem}

We consider the evolution of a thin liquid film having spatially heterogeneous viscosity.  The film lies on a flat plane and spreads under the action of surface tension alone.  The liquid wets the plane {(with zero equilibrium contact angle)} and satisfies the no-slip condition at its lower surface; its upper surface is free of external stress.  The liquid is assumed to have Newtonian rheology but contains a chemical species, which is transported passively, such that the liquid's viscosity is linearly proportional to the chemical concentration (the surface tension being unaffected).  Provided the film is sufficiently thin, lubrication theory can be used to derive a nonlinear evolution equation for the film thickness $H(X,T)$, as a function of distance $X$ along the plane and time $T$.  Molecular diffusion is assumed sufficiently strong to suppress transverse but not axial concentration gradients of the chemical species, so that its cross-sectionally averaged concentration, and thus the cross-sectionally averaged solute field $\overline{M}(X,T)$ (which for convenience we will call the viscosity field), are transported by the cross-sectionally averaged fluid velocity $\overline{U}(X,T)$.  As demonstrated in Appendix {(a)}, these equations (in a planar geometry) may be expressed {in dimensionless form} as
\begin{subequations}
\label{PDE0}
\begin{align}
&H_T + \left(\overline{U}H\right)_X  = 0,\quad \overline{U}=\frac{H^2}{3\overline{M}}H_{XXX},\\
&{\overline{M}}_T  + \overline{U}\, {\overline{M}}_X = \frac{1}{Pe}\frac{(H{\overline{M}}_X)_X}{H}.
\end{align}
\end{subequations}
The evolution equation for $H$ describes how fluid is transported by surface-tension-induced pressure gradients associated with gradients of interfacial curvature, at a rate modulated by the local viscosity field; this field is transported by bulk advection and can spread along the film via molecular diffusion.  The P\'{e}clet number $Pe$, measuring the strength of advection to diffusion, is chosen to be sufficiently large for axial diffusion to appear only as a weak singular effect.  In the absence of diffusion (\ref{PDE0}b) can be expressed in conservative form for the transported variable $H\overline{M}$, which represents the amount of solute per unit length of film. 

To illustrate the impact of heterogeneous viscosity we consider spreading of a droplet sitting on a precursor film.  The drop has an initial parabolic profile with (dimensional) height $h_0$ and half-width $l_0$, from which we define an aspect ratio $\epsilon=h_0/l_0\ll 1$; the precursor film surrounding the drop has thickness $\eta h_0$ where $\eta \ll1$. We do not attempt to model the impact of the drop with the film or subsequent mixing of material.  The initial condition on $H$ is simply
\begin{align}\label{ICH}
H(X,0) = 
\begin{cases}
\eta& (|X|>1), \\
\eta+1 - X^2 & (|X|\leq 1).
\end{cases}
\end{align}
The initial viscosity field $\overline{M}(X, 0)$ is represented as a random field $\mathcal{M}(X, \omega)$, where $\omega$ is an event in an underlying probability space. For fixed $X$,  $\mathcal{M}$ is a random variable; for an outcome $\omega$, $\mathcal{M}$ is a function of $X$ that we call the sample associated with $\omega$. We assume  $ \mathcal{M} = \exp(\mathcal{G}(X,\omega))$, where $\mathcal{G}(X,\omega)$ is a Gaussian random field with zero mean and stationary covariance 
\begin{align}
k_\mathcal{G}(X, X') = \sigma^2\exp\left(-\frac{1}{2}\left(\frac{X-X'}{l}\right)^2\right).
\label{eq:covar}
\end{align}
Here $\sigma^2$ is the variance of the Gaussian random field and $l$ the correlation length of the initial viscosity distribution.  The squared exponential covariance function (\ref{eq:covar}) yields smooth samples of the Gaussian random field, facilitating numerical simulations.  For convenience, we do not label variables $H$, $\overline{M}$, \hbox{etc.} with $\omega$ although this will be implicit.  We wish to establish how the uncertainties in $\mathcal{M}$, represented by $\sigma^2$ and $l$, propagate through (\ref{PDE0}), when the precursor film is vanishingly thin ($\eta \ll 1$) and diffusion is weak ($1\ll Pe\ll \epsilon^{-1}$).

To close the problem we impose no-flux conditions at $\vert X\vert=L$ for some $L\gg 1$, ensuring that the film sufficiently far from the drop remains undisturbed as the drop spreads. To perform numerical simulation, we draw a sample of $\mathcal{M}$ (constructed using a Karhunen--Lo\'eve decomposition, see {Appendix (b)}); using this as an initial condition for $\overline{M}(X,0)$ we solve (\ref{PDE0}) numerically with the method of lines using fourth-order spatial differences.  We collect results from multiple runs to compute statistics (such as mean and variance) of quantities of interest, which we compare to predictions of asymptotic analysis.  

{Results of simulations are presented in Section 3 and in Figures 1-4.  These figures also include approximations from a low-order model, derived in Section 4 below.  We restrict attention to times over which the drop remains significantly thicker than the precursor film.}

\section{Simulations}

\begin{figure}
\begin{center}
\includegraphics[width=5in]{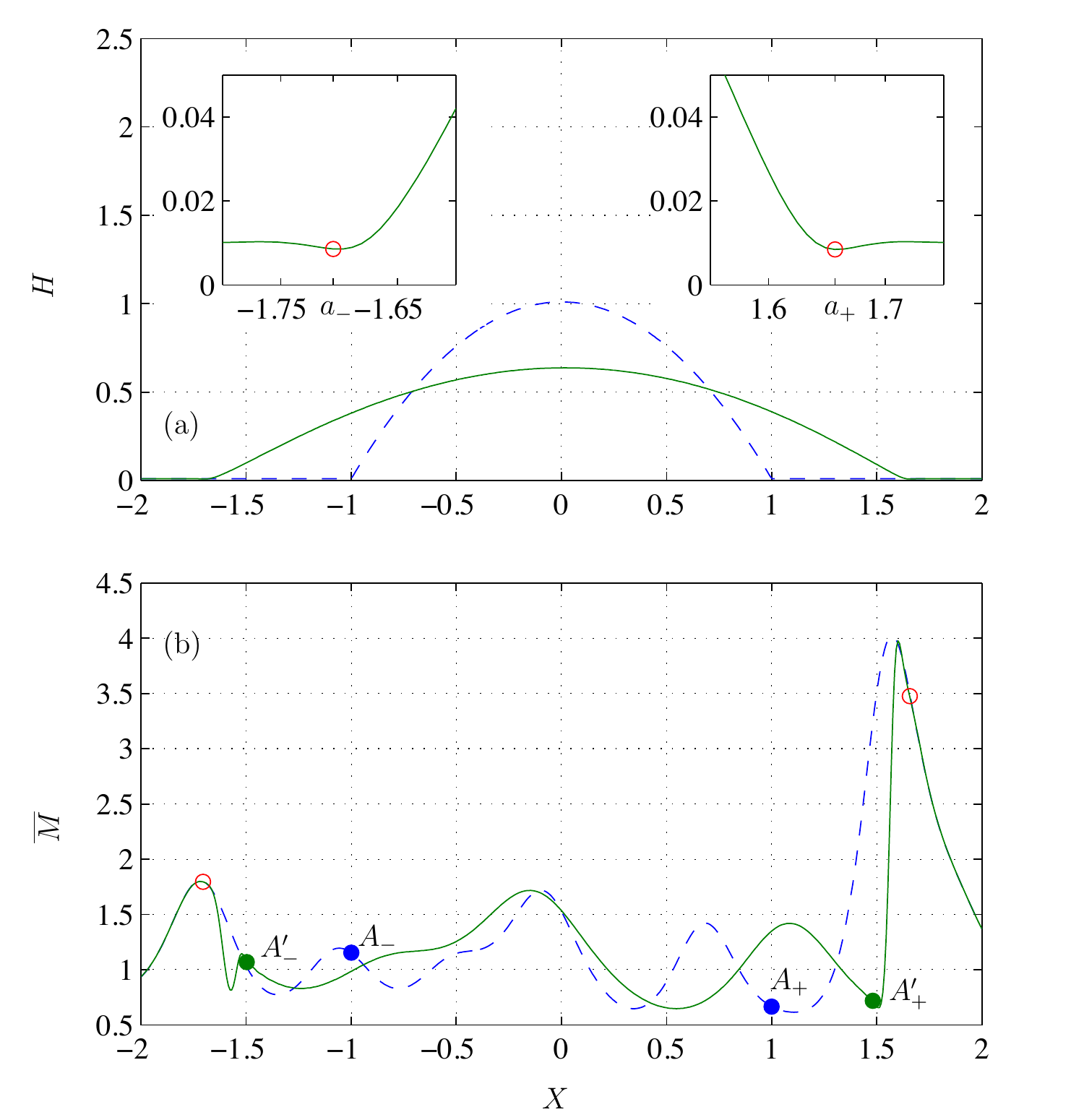}
\caption{A realisation of (\ref{PDE0}) with $\sigma = 0.5$, $l = 0.2$, $Pe = 10^{5}$, {$L=2$} and $\eta = 10^{-2}$, showing (a) $H$ and (b) $\overline{M}$ at $T = 0$ (dashed) and $T = 10$ (solid); open circles in (a,b) denote $X=a_{\pm}$ where the film thickness has a primary minimum; points marked $A_{\pm}$ in (b) show the initial drop edge $X=\pm 1$; points marked $A_{\pm}'$ show locations where $\overline{U}_X=0$ at $T=10$, distinguishing regions of expansion ($\overline{U}_X>0$ in $A_{-}'<X<A_{+}'$) from compression.}
\label{PDESE}
\end{center}
\end{figure}
Figure~\ref{PDESE} shows an example of drop spreading given a sample of  $\mathcal{M}$, for which the correlation length $l$ is shorter than the drop width and the variance $\sigma$ is sufficiently large to ensure that variations in the film's viscosity span an order of magnitude. The drop retains a parabolic profile as it spreads.  Insets near each contact line (Figure~\ref{PDESE}a) show a characteristic dimple in the film thickness where the drop connects to the precursor film.  We use the local minimum to identify the contact-line locations, defining $X = a_\pm(T)$ to be the locations at which $H$ reaches its first minimum as $X$ increases (decreases) from the drop centre, where  $a_-(T) < a_+(T)$.  We use these variables to characterise the drop width $\mathcal{W}(T)$ and lateral displacement of its mid-point $\mathcal{C}(T)$ as the drop spreads, defined by
\begin{align}
\mathcal{W}(T)=a_+ - a_-,\quad
\mathcal{C}(T)=\tfrac{1}{2}(a_+ +a_-).
\end{align}
Because the initial viscosity distribution is heterogeneous (Figure~\ref{PDESE}b), the two contact lines travel at slightly different speeds: in this example the left-hand contact line has travelled a little further than the right-hand contact line ($\vert a_-(10)\vert>a_+(10)$); the region of high viscosity near $X=1.6$ appears to restrain the motion of the right-hand contact line.  

\begin{figure}
\begin{center}
\includegraphics[width=5in]{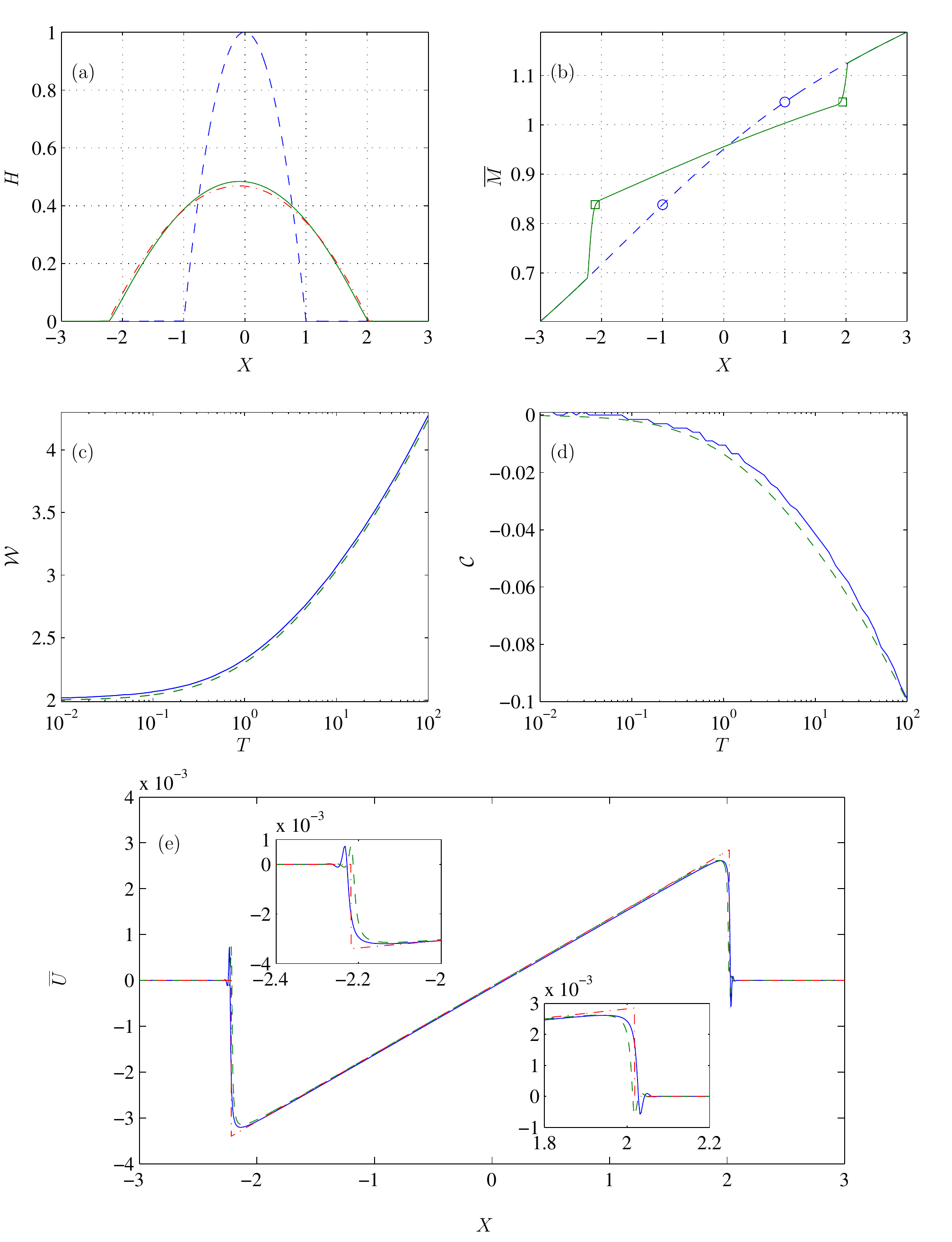}
\caption{A realisation of (\ref{PDE0}) with $\sigma = 0.5$, $l = 5$, $Pe = 10^{5}$ and $\eta = 10^{-3}$, and its asymptotic approximation. (a) $H$ at $T = 0$ (dashed) and $T = 100$ (solid) calculated from (\ref{PDE0}) and $H$ at $T = 100$ from leading-order outer asymptotic solution (\ref{OutLeadingSol}) (dash-dotted); (b) $\overline{M}$ at $T = 0$ (dashed) and $T = 100$ (solid) calculated from (\ref{PDE0}), where circles show initial values of $M$ at $X =  -1$ and $X = 1$, that is $M_+ = 1.046$ and $M_- = 0.8382$, and squares show the position of $M_{\pm}$ at $T = 100$ convected by $\overline{U}$; (c) drop width $\mathcal{W}$ calculated from (\ref{PDE0}) (solid) and from {the low-order model} (\ref{ODE}, \ref{eq:alpha}) (dashed); (d) drop centre $\mathcal{C}$ calculated from (\ref{PDE0}) (solid) and from (\ref{ODE}, \ref{eq:alpha}) (dashed); (e) bulk velocity $\overline{U}$ (solid) {and its asymptotic approximations} $\overline{U}_{\mathrm{out}}$ (\ref{OutVelocity}) (dashed) and $\overline{U}_{\mathrm{com}}$ (\ref{ComVelocity}) (dot-dashed), at $T = 100$, with insets showing close-ups near each contact line.}
\label{PDEODE}
\end{center}
\end{figure}
The simulation in Figure~\ref{PDEODE} demonstrates how the drop behaves when the correlation length $l$ is large compared to the drop width.  The viscosity is larger on the right-hand side of the drop, leading to slight leftward displacement of the drop centre as it spreads (Figure~\ref{PDEODE}d).  As in the majority of cases investigated, the bulk velocity of the spreading flow $\overline{U}$ (Figure~\ref{PDEODE}e) is approximately linear beneath the drop, falling abruptly to zero (with small flow reversal) near each contact line.  

\begin{figure}
\begin{center}
\includegraphics[width=5in]{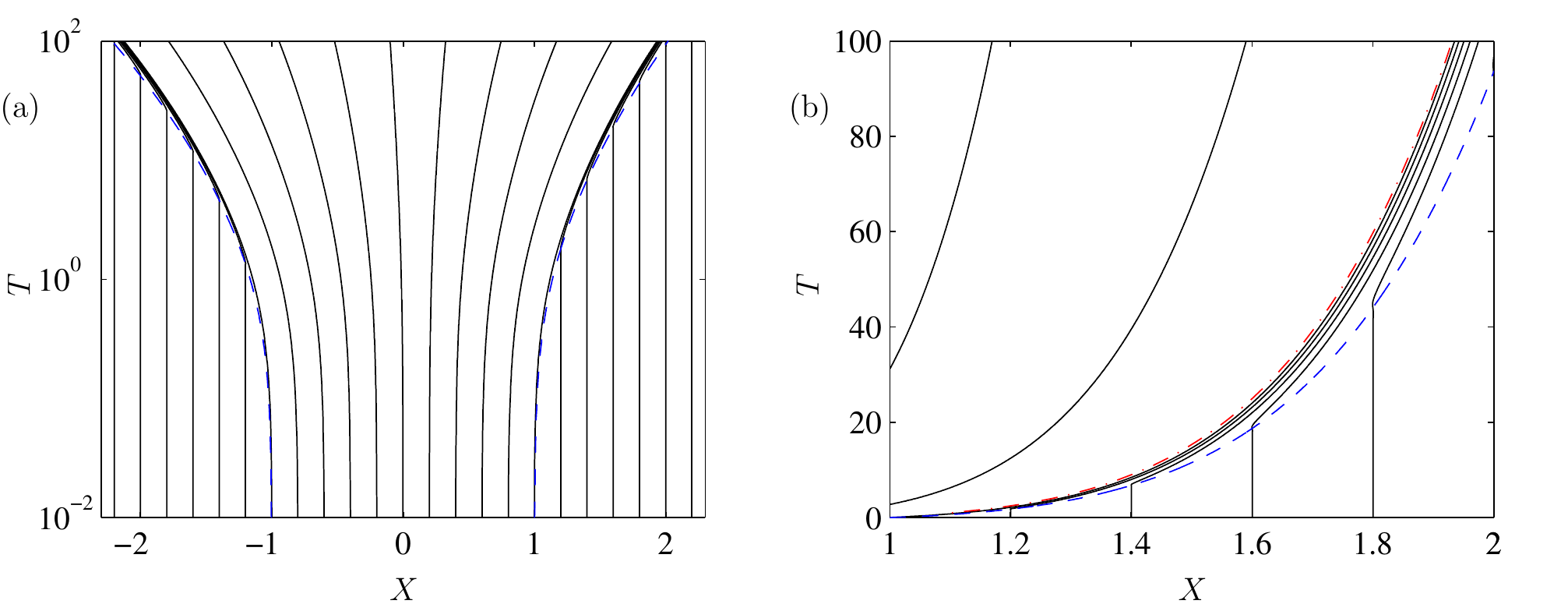}
\caption{(a) Contact-line locations $X=a_{\pm}(T)$  (dashed) and characteristics (solid) calculated from $\text{d}X/\text{d}T = \overline{U}$ for the flow presented in Figure~\ref{PDEODE}, using $\overline{U}=\overline{U}_{\mathrm{com}}$.  (b) Near the right-hand contact line, characteristics cross $X=a_+(T)$ and accumulate in a narrow region behind it.  The dot-dashed line indicates the characteristic along which $\overline{U}_X=0$, lying in the overlap between the inner and outer regions.  {$T$ is plotted on a log scale in (a) and a linear scale in (b).}}
\label{fig:charX}
\end{center}
\end{figure}
Transport of the $\overline{M}$ field may be understood by considering (\ref{PDE0}b) in the absence of diffusion, which may be expressed in terms of characteristics as
\begin{align}
\frac{\mathrm{d}\overline{M}}{\mathrm{d}T}=0 \quad \mathrm{on}\quad
\frac{\mathrm{d}X}{\mathrm{d}T}=\overline{U}.
\label{eq:char}
\end{align}
Thus the linear stretching flow beneath the spreading drop ($\overline{U}_X>0$, Figure~\ref{PDEODE}e) stretches the $\overline{M}$ field laterally without changing its magnitude.  This is illustrated in Figure~\ref{PDESE}(b), where the points $A_{\pm}'$ bound the region in which $\overline{U}_X>0$.  Ahead of the drop the $\overline{M}$ field is undisturbed, while near the contact line, where the flow is strongly compressive ($\overline{U}_X<0$, see insets in Figure~\ref{PDEODE}e), the $\overline{M}$ field steepens.  This compression is evident from the distributions in Figures~\ref{PDESE}(b) and \ref{PDEODE}(b).  In Figure~\ref{PDEODE}(b), symbols mark locations at which $\overline{M}(X,T)=\overline{M}(\pm 1,0)$, demarcating the boundary between stretching and compression of the $\overline{M}$ field.  This is further demonstrated by the pattern of characteristics, which shows uniform stretching of the concentration field beneath the drop (Figure~\ref{fig:charX}a), with crowding of characteristics near the contact line (Figure~\ref{fig:charX}b), leading to rapid variation of $\overline{M}$ in this region.  Weak axial diffusion can be expected to suppress such gradients over long times.

\begin{figure}
\begin{center}
\includegraphics[width=5in]{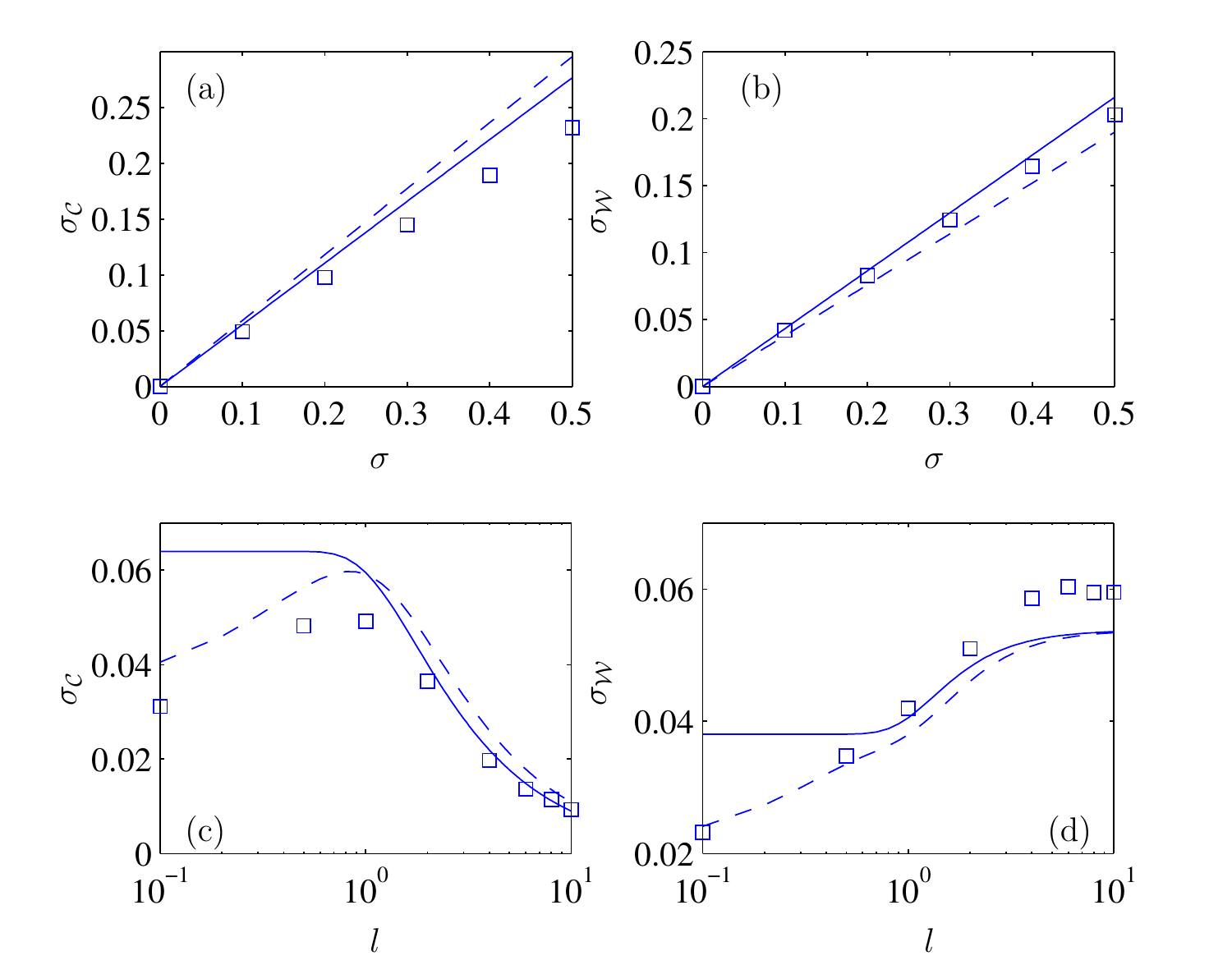}
\caption{The effect of variance $\sigma$ on the standard deviation (a,b) of drop centre $\mathcal{C}$ and width $\mathcal{W}$ respectively at $T= 100$, for $l = 1$, $\eta = 10^{-3}$ and $Pe = 10^{5}$. Squares represent results from the PDE model (\ref{PDE0}); 
solid lines represent results from the weak disorder approximation  (\ref{AsymMeanVar}) and dashed lines represent results from the linearised wedge model (\ref{AsymMeanVarWedge}). (c,d) show corresponding dependence on {the} correlation length $l$ for $\sigma=0.1$. }
\label{EOV}
\end{center}
\end{figure}

Figure~\ref{EOV} presents statistics describing drop spreading over multiple realisations of the initial viscosity field.  We use 1000 samples to estimate the standard deviation of the drop centre and width ($\sigma_{\mathcal{C}}$, $\sigma_{\mathcal{W}}$) at $T= 100$ and assess the dependence on the variance $\sigma^2$ and correlation length $l$ of the initial viscosity field; { the means of $\mathcal{C}$ and ${\mathcal{W}}$ do not show appreciable dependence on $\sigma$ or $\ell$ in this example}.  For the present we focus on the square symbols, denoting predictions from simulations of (\ref{PDE0}).   Figure~\ref{EOV}(a,b) shows that, as $\sigma$ increases, $\sigma_{\mathcal{C}}$ and $\sigma_{\mathcal{W}}$ increase.  (Simulations for larger $\sigma$ were limited by the difficulty of resolving very large viscosity gradients that accumulated in the contact-line region, for the chosen value of $Pe$.)  The standard deviations show noticeable dependence on {the} correlation length (Figure~\ref{EOV}c,d): for small $l$, the viscosities at the left and right contact lines are uncorrelated, whereas they become increasingly similar as $l$ increases.  Consequently, fluctuations in drop width increase in magnitude as $l$ increases (if one contact line is, say, hindered, then the other is also likely to be), while there is less tendency for the drop to drift sideways (the mean drift remains very close to zero).  This behaviour is consistent with studies of drops spreading on random surfaces \cite{savva2010two}.  As $l$ becomes very small, $\sigma_\mathcal{C}$ falls; this reflects the effects of axial diffusion in simulations suppressing sharp gradients in the solute field.  We seek to quantify the dependence of $\sigma_{\mathcal{C}}$ and $\sigma_{\mathcal{W}}$ on $\sigma$ and $l$ using an asymptotic model below.

\section{Derivation of a low-order model}

With $\eta\ll 1$, the drop motion is slow and is dominated by the flow in the neighbourhood of the contact lines.   We now investigate the impact of readjustment of the viscosity field on this motion, {initially} neglecting the influence of axial diffusion in (\ref{PDE0}).  We divide the flow into an outer region in which the drop adopts an equilibrium shape to leading order, with narrow regions at each contact line (illustrated by insets to Figure~\ref{PDESE}(a)) governed by a modified form of the Landau--Levich equation.  While the overall structure of the flow follows the uniform-viscosity case \cite{hocking1983,savva2009,eggers2015singularities}, we seek to identify how variations of film properties modify the drop spreading rate.   We follow previous authors \cite{savva2009, sibley2015asymptotics} in matching the cube of the interface slope between inner and outer regions, rather than invoking an intermediate region.  Formally, we assume that $\sigma$ and $l$ are $O(1)$ as $\eta\rightarrow 0$, ensuring that the correlation length of the viscosity field exceeds the width of the contact-line regions.
 
\subsection{Outer region}

For $X>a_+$ and $X<a_-$, away from the contact lines, $H = \eta$, $\overline{M} = \overline{M}(X, 0)$ and $\overline{U} = 0$.  Within the drop, with $a_-(T) < X < a_+(T)$, we seek a solution of (\ref{PDE0}a) subject to 
\begin{align}
&\lim_{X \to a_\pm \mp}H = 0, \quad \int_{a_-}^{a_+} H dX = \mathcal{V},
\label{eq:bc1}
\end{align}
where $\mathcal{V}=4/3$ is the volume of the droplet {(and the $\mp$ symbol here denotes a one-sided limit)}.  {Assuming the drop shape to be quasi-static,} we write
\begin{align}
X =   \mathcal{C}+\tfrac{1}{2}\mathcal{W} Y, \quad H(X, T) = G(Y; a_+, a_-), \quad \overline{M}(X,T)=N(Y;a_+,a_-)
\end{align}
which allows us to write (\ref{PDE0}a) as
\begin{align}
G_{a_+} \dot{a}_+ + G_{a_-} \dot{a}_- - \frac{\dot{a}_+(1 + Y) + \dot{a}_-(1 - Y)}{\mathcal{W}} G_Y + \frac{1}{3}\left(\frac{2}{\mathcal{W}}\right)^4 \left(\frac{G^3 G_{YYY}}{N}\right)_Y = 0,
\label{eq:G}
\end{align}
where $\dot{a}_+ \equiv {\mathrm{d}a_{+}/\mathrm{d}T} >0$ and $\dot{a}_- \equiv {\mathrm{d}a_{-}/\mathrm{d}T} <0$.  Assuming $|\dot{a}_{\pm}|  \ll 1$ (we will see below that $\dot{a}_{\pm}$ is approximately $O(1/\log(1/\eta))$ when $T=O(1)$), we expand $G$ as $G = G_0 + G_1 + \cdots$ where $G_0$ is the quasi-static solution
\begin{align}\label{OutLeadingSol}
G_0 = \tfrac{3}{2} \mathcal{V} \left(1-Y^2\right)/ \mathcal{W}
\end{align}
and $G_1$ is {linear in} $\dot{a}_{\pm}$, satisfying
\begin{align}\label{OutSecondEq}
G_{0a_+} \dot{a}_+ + G_{0a_-} \dot{a}_- - \frac{\dot{a}_+(1 + Y) + \dot{a}_-(1 - Y)}{\mathcal{W}} G_{0Y} + \frac{1}{3}\left(\frac{2}{\mathcal{W}}\right)^4 \left(\frac{G_0^3 G_{1YYY}}{N}\right)_Y = 0,
\end{align}
which we seek to solve subject to
\begin{align}\label{OutSecondBC}
&\lim_{Y \to \pm 1\mp}G_1 = 0, \quad \int_{-1}^{1} G_1 dY = 0.
\end{align}
Substituting (\ref{OutLeadingSol}) into (\ref{OutSecondEq}) and integrating  (\ref{OutSecondEq}) once with respect to $Y$ yields
\begin{align}\label{OutSecondEq2}
&G_{1YYY} =  \frac{N \mathcal{W}^5 \big((\dot{a}_+ -\dot{a}_-) Y+\dot{a}_++\dot{a}_-\big)}{12 \mathcal{V}^2 \left(1 - Y^2\right)^2}.
\end{align}
The bulk fluid velocity in the outer region is therefore
\begin{align}\label{OutVelocity}
\overline{U}_{\mathrm{out}} &= \frac{H^2H_{XXX}}{3\overline{M}} = \frac{8(G_0^2G_{1YYY} + \cdots)\mathcal{H}(a_+-X)\mathcal{H}(X-a_-)}{3N \mathcal{W}^3} \nonumber \\
&= \frac{ (X-a_-)\dot{a}_+ +(a_+-X)\dot{a}_-}{\mathcal{W}}\mathcal{H}(a_+-X)\mathcal{H}(X-a_-) + \cdots,
\end{align}
where $\mathcal{H}(\cdot)$ is Heaviside function.  The linear stretching flow is illustrated in Figure~\ref{PDEODE}(c).  Thus $\overline{U}_{\mathrm{out}}=\dot{C}+\tfrac{1}{2}\dot{W}Y\equiv \mathrm{d}X/\mathrm{d}T$, implying that $\overline{M}=N(Y)$ exactly satisfies (\ref{eq:char}).  The viscosity field beneath the bulk of the drop is stretched linearly, as illustrated in Figures \ref{PDESE}(b) and \ref{PDEODE}(b).  It is therefore reasonable to identify $M_\pm(T) =  \lim_{Y \to \pm 1} N= \overline{M}(a_\pm(0+),0)$, where $0+$ denotes the early time at which the asymptotic spreading structure is established (which we take here to be $T=0$); the value of $M_{\pm}$ may be affected by axial diffusion in practice. 

$G$ is singular as $Y \rightarrow \pm1$ and (\ref{OutSecondEq2}) shows its asymptotic behaviour to be
\begin{align}
\label{OutSecondAsy}
G_{1YYY} = \frac{ M_\pm\dot{a}_\pm \mathcal{W}^5}{24 \mathcal{V}^2 (1{\mp}Y)^2} + O((1 {\mp} Y)^{-1}), \qquad (Y \rightarrow  \pm1\mp).
\end{align}
Integrating (\ref{OutSecondAsy}) twice with respect to $Y$ gives
\begin{align}
\label{OutSecondIntegral}
G_{1Y}  \sim -\frac{ M_\pm \dot{a}_\pm \mathcal{W}^5}{24 \mathcal{V}^2 }\Big[\ln (1 {\mp} Y) + \zeta_{\pm} +{O((1\mp Y)\ln(1\mp Y))} \Big] , \qquad (Y \rightarrow \pm1\mp).
\end{align}
Finding the constants $\zeta_\pm$ requires use of conditions (\ref{OutSecondBC}), which cannot be carried out analytically for arbitrary $N(Y)$.  However, {following \cite{savva2009}} and multiplying  (\ref{OutSecondEq2}) by $(1-Y)(1+Y)^2$ and integrating by parts with respect to $Y$ from $-1+\epsilon_-$ to $1-\epsilon_+$ ($\epsilon_-$  and $\epsilon_+$ are small and positive), we have
\begin{gather}
\big[(1-Y)(1+Y)^2 G_{1YY} - (1+Y)(1-3Y) G_{1Y} - 2(1+3Y) G_1\big]_{-1+\epsilon_-}^{1-\epsilon_+} + 6\int_{-1+\epsilon_-}^{1-\epsilon_+} G_1 \text{d} Y \nonumber \\
=\frac{\mathcal{W}^5}{12\mathcal{V}^2}\left[(\dot{a}_+ - \dot{a}_-)\int_{-1+\epsilon_-}^{1-\epsilon_+} \frac{N Y}{1-Y} \text{d} Y + (\dot{a}_+ + \dot{a}_-)\int_{-1+\epsilon_-}^{1-\epsilon_+} \frac{N}{1-Y} \text{d} Y \right].
\label{eq:MultiplyandIntegrateX}
\end{gather}
Utilising (\ref{OutSecondBC}, \ref{OutSecondIntegral}) and noticing that %
{
\begin{subequations}
\begin{align}
\int_{-1+\epsilon_-}^{1-\epsilon_+} \frac{N}{1-Y} \text{d} Y &\approx M_-(\ln 2 - M_+\ln \epsilon_+) + \int_{-1+\epsilon_-}^{1-\epsilon_+} N_Y \ln(1-Y)\,\mathrm{d}Y, \\
\int_{-1+\epsilon_-}^{1-\epsilon_+} \frac{N Y}{1-Y} \text{d} Y &\approx M_-(\ln 2 -1) - M_+(1+\ln \epsilon_+) + \int_{-1+\epsilon_-}^{1-\epsilon_+} N_Y(Y+\ln(1-Y))\,\mathrm{d}Y 
\end{align}
\end{subequations}
}
(provided $N$ is sufficiently smoothly varying), we simplify (\ref{eq:MultiplyandIntegrateX}) to find, {as $\epsilon_\pm \to 0$, 
\begin{equation}
\zeta_\pm = 1+\frac{\mathcal{I}_1- \mathcal{I}_2^\pm - M_\mp \ln 2}{M_\pm}  - \frac{\dot{a}_\mp}{\dot{a}_\pm} \frac{\mathcal{I}_1}{M_\pm},
\end{equation}
}
where the result for $\zeta_-$ follows analogously from multiplying  (\ref{OutSecondEq2}) by $(1+Y)(1-Y)^2$.  {Here
\begin{equation}
\mathcal{I}_1\equiv \tfrac{1}{2}\int_{-1}^1 N\,\mathrm{d}Y,\quad
\mathcal{I}_2^\pm\equiv \int_{-1}^1 N_Y \ln(1\mp Y)\,\mathrm{d}Y, \quad
N(Y)\equiv \overline{M} \left(\mathcal{C}(0)+\tfrac{1}{2}\mathcal{W}(0)Y,0\right).
\end{equation}
}

The asymptotic behaviour of $H_X$ (the inner limit of the outer solution) is obtained as
\begin{equation}
H_X = \frac{2}{\mathcal{W}} G_Y \sim \frac{2}{\mathcal{W}} \left(\mp\frac{3 \mathcal{V}}{\mathcal{W}}  -\frac{ M_\pm\dot{a}_\pm \mathcal{W}^5}{24 \mathcal{V}^2 }\Big[\ln (1 \mp Y) + \zeta_\pm \Big] \right), \quad (Y \rightarrow \pm1\mp),
\end{equation}
{(recalling \cite{hocking1983, savva2009})} so the cube of the slope may be written in original variables as
\begin{align}
\label{CubeOuta}
H_X^3 = \mp \frac{216\mathcal{V}^3}{\mathcal{W}^6} - 9 M_\pm\dot{a}_\pm 
\left[\ln \{\pm(a_\pm-X)\} + \ln \frac{2}{\mathcal{W}} +\zeta_\pm \right], \qquad  (X \rightarrow a_\pm \mp).
\end{align}

\subsection{Inner region}

For convenience we restrict attention initially to the inner region at the right-hand contact line.  Here
we stretch $X$ and enlarge $H$ using
\begin{align}
& X = a_+ + {\eta}{(3 \dot{a}_+)^{-1/3}} \tilde{X}, \quad H(X,T) = \eta \tilde{H}(\tilde{X}),\quad \overline{M}(X,T)=\tilde{M}(\tilde{X}),
\end{align}
so that (\ref{PDE0}a) becomes, for $\eta \ll 1$, 
\begin{align}\label{InLeading3}
&\tilde{M}(1 - \tilde{H}) + {\tilde{H}^3 \tilde{H}_{\tilde{X}\tilde{X}\tilde{X}}} = 0
\end{align}
after integrating once and imposing $\tilde{H}\rightarrow 1$ as $\bar{X}\rightarrow \infty$.  Equation (\ref{InLeading3}) is a generalised Landau--Levich equation, modified by the variable viscosity field $\tilde{M}$. We see from Figure~\ref{fig:charX} that characteristics crowd into the contact-line region, reflecting the compression of the viscosity field that is encountered by the advancing contact line.  While the compressed field drifts slowly from the front to the back of the inner region, we assume for the time being that the $\tilde{M}$ field is quasi-steady; we return to its slow evolution later on. 

Matching to the outer region requires $\tilde{M}\rightarrow M_+$ for $\tilde{X}\rightarrow -\infty$ (in the overlap between the inner and outer regions), and $\lim_{\tilde{X} \to \infty}\tilde M = M^{+} \equiv \overline{M}(a_+(T)+,0)$, representing the contact line advancing over the unperturbed viscosity field. These far-field boundary conditions on $M$ help us to construct far-field asymptotic solutions of (\ref{InLeading3}).  One boundary condition can be simplified by writing $\hat{X} =  (M^{+})^{1/3}\tilde {X}$, $\hat{H}(\hat{X})=\tilde{H}(\tilde{X})$, $\hat{M}(\hat{X}) = \tilde{M}(\tilde{X})/M^{+}$, so that (\ref{InLeading3}) becomes
\begin{align}
\hat{M}(1 - \hat{H}) + {\hat{H}^3 \hat{H}_{\hat{X}\hat{X}\hat{X}}} = 0.
\label{InLeading4}
\end{align}
The corresponding boundary conditions are
\begin{align}\label{InLeading4BC}
& \lim_{\hat{X} \to -\infty} \tilde{H}_{\hat{X}\hat{X}} = 0 \quad\mathrm{where}\quad \hat{M} = r_M\equiv M_+/M^{+},\quad 
\lim_{\hat{X} \to \infty} \tilde{H} = 1 \quad\mathrm{where}\quad \hat{M} = 1.
\end{align}
We are also at liberty to place the primary minimum of the solution at the origin ($\tilde{H}_{\hat{X}}(0) = 0$) in computed solutions. 

Denoting $\bar{\bar{X}} = r_M^{1/3}\hat{X}$, (\ref{InLeading4}) has an asymptotic solution of the form
\begin{align}\label{asy_sol_n_inf}
\hat{H} = -\bar{\bar{X}} \phi(\bar{\bar{X}})^{1/3}\left(3^{1/3} + \alpha_1 \phi(\bar{\bar{X}})^{-1} + O(\phi(\bar{\bar{X}})^{-2})\right) , \quad (\bar{\bar{X}} \rightarrow -\infty),
\end{align}
where $\alpha_1$ is a constant dependent on the whole $\hat{M}$ field in the inner region and $\phi(\bar{\bar{X}}) \equiv \ln (-\bar{\bar{X}})$. As $\hat{X} \to +\infty$, (\ref{InLeading4}) has an asymptotic solution of the form
\begin{align}\label{asy_sol_p_inf}
\tilde{H} = &1 + \alpha_2 \exp\left(-\frac{\hat{X}}{2}\right)\cos \left(\frac{\sqrt{3} \hat{X}}{2}\right) + O\left(\exp\left(-\hat{X}\right)\right), \quad (\hat{X} \rightarrow +\infty),
\end{align}
where $\alpha_2$ is also a constant. The solution (\ref{asy_sol_p_inf}) represents a one-parameter family of solutions of (\ref{InLeading4}); only one member of that family satisfies (\ref{asy_sol_n_inf}). We solve (\ref{InLeading4}) numerically to determine $\alpha_2$, and hence $\alpha_1$.    Shooting towards $\hat{X} \to -\infty$, we seek the solution satisfying $\hat{X} \tilde{H}_{\hat{X}}^2\tilde{H}_{\hat{X}\hat{X}} = -r_M$ at the left of the domain, in accordance with the asymptotic solution (\ref{asy_sol_n_inf}).   We evaluate $\alpha_1$ by solving $\tilde{H}_{\hat{X}}^3 + 3 r_M(\ln(-r_M^{1/3}\hat{X})+3^{2/3}\alpha_1 +1) = 0$ at the domain boundary.  

\begin{figure}
\begin{center}
\begin{overpic}[width=5in]{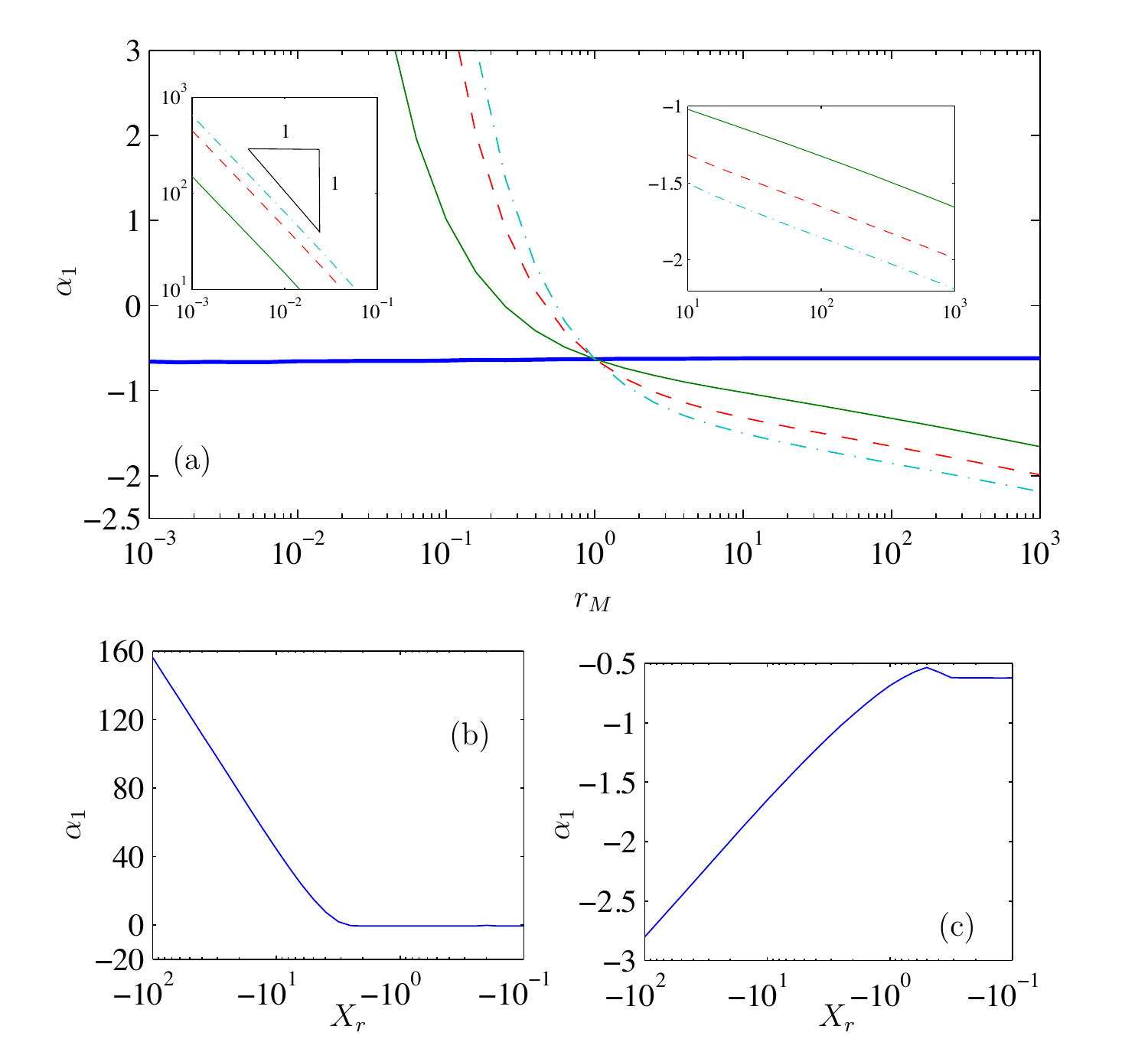}\end{overpic}
\caption{(a) $\alpha_1$ versus $r_M$ with $X_r =0$ (horizontal line),  $-5$ (solid), $-10$ (dashed) and $-15$ (dot-dashed) and  $\alpha_1$ versus $-X_r$ with (b) $r_M = 0.01$ and (c) $100$, from numerical solutions of (\ref{InLeading4}, \ref{InLeading4BC}) with step viscosity (\ref{stepM}) and  $\hat{X}_L = 10^9$ and $\hat{X}_R = 10$.}
\label{alpha_1}
\end{center}
\end{figure}

When $\hat{M}\equiv 1$, for example, we find that $\alpha_1\approx -0.63\equiv A_1$, say, in accordance with prior studies { (Peng \hbox{et al.} \cite{peng2015}, for example, report a value equivalent to $-0.61$ using a three-term expansion over an unspecified domain)}.   We now consider how $\alpha_1$ is influenced by spatial variations of the $\tilde{M}$ field.  Because the inner-region flow is compressive, a region of high or low viscosity encountered by an advancing contact line will typically manifest as a steep ramp in the $\tilde{M}$ field; the ramp will slowly propagate from the front to the rear of the inner region. To mimic this situation, we chose 
\begin{align}\label{stepM}
\hat{M} = \begin{cases}
r_M, & (\hat{X} < \hat{X}_r), \\
1, & (\hat{X}\geq \hat{X}_r),
\end{cases}
\end{align}
for some $\hat{X}_r$, imposing continuity conditions in $\hat{H}$, $\hat{H}_{\hat{X}}$, $\hat{H}_{\hat{X}\hat{X}}$ across $\hat{X}_r$, which we solved on a long domain $[-\hat{X}_L, \hat X_R]$.    Computed values of $\alpha_1$ are illustrated in Figure~\ref{alpha_1}.  For values of $r_M$ close to unity, the approximation $\alpha_1\approx A_1$ is robust, particularly when $\hat{X}_r>0$.  However there is a striking difference between cases in which $r_M\ll 1$ (representing a drop spreading into a high-viscosity region) and $r_M\gg 1$ (a drop encountering a low-viscosity region): $\alpha_1$ becomes large and positive in the former case (with $\alpha_1\propto 1/r_M$ for fixed $X_r\leq 0$) but remains relatively small and negative in the latter.   The jump in viscosity therefore has greatest influence on the magnitude of $\alpha_1$ when the jump extends into regions where the film is thicker and when the contact line is encountering a region of elevated viscosity.   We explore the consequences of these variations below.

\subsection{Matching}

Written in outer variables, the outer limit of the inner problem (\ref{asy_sol_n_inf}) is
\begin{align}
H_X^3 = 3 \dot{a}_+ M_+ \tilde{H}_{\bar{\bar{X}}}^3 = 3 \dot{a}_+ M_+ \left(-3\ln (- \bar{\bar{X}}) - 3 - 3^{5/3}\alpha_1 + O(\ln (- \bar{\bar{X}})^{-1}) \right)  \nonumber 
\end{align}
so that in the overlap regions $\eta\vert \dot{a}_\pm\vert^{-1/3} \ll \vert a_\pm -X\vert  \ll 1$,
\begin{align}
\label{CubeIna}
H_X^3 \approx 9 \dot{a}_\pm M_\pm \left(\ln \eta -\ln\vert a_\pm -X\vert  - \tfrac{1}{3} \ln\vert 3\dot{a}_\pm M_\pm \vert -1 - 3^{2/3}\alpha_{\pm}\right),
\end{align}
where $\alpha_{\pm}$ denotes the values of $\alpha_1$ at each contact line.
Matching (\ref{CubeOuta}) with (\ref{CubeIna}) 
gives the coupled ODEs
{
\begin{subequations}
\label{ODE}
\begin{align}
\left[ M_{\pm} \left(\mathcal{J}_\pm  + \tfrac{1}{3} \ln \vert 3\dot{a}_\pm M_\pm\vert \right) -\mathcal{I}_1 - (M_\pm-M_\mp)\ln 2+\mathcal{I}_2^\pm\right]\dot{a}_{\pm} + \mathcal{I}_1 \dot{a}_\mp=\pm {24 \mathcal{V}^3}/{\mathcal{W}^6}
\end{align}
where
\begin{align}
\mathcal{J}_{\pm} \equiv 3^{2/3} \alpha_{\pm}+\ln\mathcal{W} -\ln\eta.
\end{align}
\end{subequations}
}
The system (\ref{ODE}) constitutes a simplified model for the slow spreading of a drop over a heterogeneous film, {and recalls similar descriptions of drop-spreading on homogeneous films \cite{hocking1992}, for which $\mathcal{I}_1=1$, $\mathcal{I}_2^\pm=0$, $M_{\pm}=1$; this limit yields a deterministic expression for $a_{\pm}=\pm a_0$, namely}
\begin{align}\label{loeq}
& \dot{a_0}
\left(-2+3^{2/3} A_1+\ln(2a_0)-\ln\eta + \tfrac{1}{3} \ln(3\dot{a_0})\right)  = \frac{3 \mathcal{V}^3}{8 a_0^6},
\end{align}
which may be written as 
$\dot a_0 a_0^6 \ln\left[ {2a_0 (3\dot a_0)^{1/3} e^{3^{2/3}A_1}}/{\eta e^2}\right]={8}/{9}$ (illustrating how the drop width grows roughly proportionally to $t^{1/7}$ in a planar geometry).

Equation~(\ref{ODE}) indicates that the contact-line speeds are coupled via hydrodynamic effects within the bulk drop, and are dependent {primarily} on the viscosities $M_{\pm}$ upstream of each contact line.  {This becomes evident on taking leading-order terms as $\eta\rightarrow 0$, when (\ref{ODE}) reduces to
\begin{equation}
\label{eq:short}
\left[ M_{\pm} \left(-\ln\eta   \right) \right]\dot{a}_{\pm} \approx \pm {24 \mathcal{V}^3}/{\mathcal{W}^6},
\end{equation}
confirming that the contact-line speed is $O(1/\ln(1/\eta))$.} The compressed viscosity field within each contact-line region influences the constants $\alpha_{\pm}$ in (\ref{ODE}b): if variations are modest, we adopt the approximation 
\begin{align}
\alpha_{\pm}=A_1\approx -0.63,\qquad M_{\pm}=\overline{M}(\pm1,0); 
\label{eq:alpha}
\end{align}
alternatively, large variations in the viscosity field are accommodated by changes in the value of $\alpha_{\pm}$ and, potentially, $M_{\pm}$.

To solve (\ref{ODE}, \ref{eq:alpha}), we transform them to a system of differential-algebraic equations by defining $\dot{a}_\pm$ as two new variables. Figure~\ref{PDEODE}(a-c) compares the asymptotic predicted drop shape, width $\mathcal{W}$ and centre $\mathcal{C}$ with simulations of the PDE system (\ref{PDE0}).  The ODE model successfully captures the lateral drift of the drop due to the gradient in the viscosity field.  Factors limiting the accuracy of the low-order model are the inclusion of diffusion in the PDE simulations; given the logarithmic (rather than algebraic) dependence on the small parameter $\eta$, { (\ref{ODE}) provides notably greater quantitative accuracy than (\ref{eq:short})}.  

The limitations of the assumption $\alpha_{\pm}=A_1$ are illustrated in Figure~\ref{PDESE}.  In this example, $M_+<M_-$ (compare the viscosities at $A_{\pm}$), leading to initial rightward drift of the drop.  However, the peak in viscosity near $X=1.6$ causes the right-hand contact line to slow and the drop to then drift left.  In this example, the viscosity field within the inner region (between $A'_+$ and the open circle in Figure~\ref{PDESE}(b)) has a ramp with magnitude $r_M\approx 0.2$.  As it propagates into the contact-line region, $\alpha_+$ can be expected to increase as indicated in Figure~\ref{alpha_1}.   The dominant terms in (\ref{ODE}) when $\eta\ll 1$ and $\alpha_+\gg 1$,
\begin{align}\label{threefactors}
\dot{a}_{+} \propto \frac{1}{M_{+}\left(\ln(1/{\eta}) + 3^{2/3}\alpha_{+}\right)},
\end{align}
indicate how passage of the region of elevated viscosity backwards into the inner region slows the advance of the contact line.

\subsection{The bulk velocity field}

To understand solute transport in more detail we use the asymptotic approximation to describe the bulk velocity field.  In the inner region, the velocity is
\begin{align}
\label{InVelocitya}
\overline{U}_{\mathrm{in}\pm}
= \frac{\tilde{H}^2 \tilde{H}_{\tilde{X}\tilde{X}\tilde{X}}}{3 \tilde{M}} 
= \dot{a}_{\pm} \left(1-
\frac{1}{\tilde{H}(\pm(X-a_\pm) (3 \vert \dot{a}_\pm\vert M_\pm)^{1/3}/\eta)}
\right).
\end{align}
Noting that $\lim_{X \to a_\pm {\mp}} U_{\mathrm{out}} = \dot{a}_\pm = \lim_{\hat{X} \to -\infty} U_{\mathrm{in}\pm }$, 
we can construct a composite approximation of $\overline{U}$ using (\ref{OutVelocity}) as
\begin{align}\label{ComVelocity}
& \overline{U}_{\mathrm{com}} = \overline{U}_{\mathrm{out}} + \overline{U}_{\mathrm{in}+} + \overline{U}_{\mathrm{in}-} - \dot{a}_+ \mathcal{H}(a_+-X)- \dot{a}_- \mathcal{H}(X - a_-).
\end{align}
This approximation is illustrated in Figure~\ref{PDEODE}(e) and is used to construct the characteristics along which $\overline{M}$ is transported in Figure~\ref{fig:charX}, using solutions of $a_{\pm}(T)$ from (\ref{ODE}, \ref{eq:alpha}) and the numerical solution of $\tilde{H}$ from (\ref{InLeading4}) with $M' = 1$.  It is clear from Figure~\ref{fig:charX}(b) that $\overline{U}_{\mathrm{com}}<\dot a_+$ in the neighbourhood of the contact line, implying that characteristics move smoothly through the inner region.  Evaluating the $X$ derivative of $\overline{U}_{\mathrm{com}}$ reveals that the velocity maximum lies a distance of order $[a_+ \eta(\dot{a}_+ M_+)^{-1/3}]^{1/2}$ (the geometric mean of the inner and outer lengthscales) behind $a_+$, placing it formally within the overlap region between the inner and outer solutions.   This defines the boundary between expansive and compressive regions (Figure~\ref{fig:charX}).  Thus, in the absence of diffusion, all the solute swept up by the contact line remains confined to a narrow zone immediately behind the contact line, within which the asymptotic inner region is confined.

\subsection{Weak disorder}

When $\sigma\ll 1$, we can use a perturbation method to quantify the variability in solutions of (\ref{ODE}, \ref{eq:alpha}) explicitly. We write the random variables $M_{\pm}$ and $a_{\pm}$ as sums of their mean and a small random perturbation
\begin{align}\label{expansion}
&M_\pm = 1 +M_{\pm 1}+\dots, \quad a_\pm = \pm a_{0} + a_{\pm 1}+\dots, 
\end{align}
{where $a_0$ satisfies (\ref{loeq})}.   We anticipate that perturbations are $O(\sigma)$ smaller than leading-order terms and 
$M_{\pm1}=\mathcal{G}(\pm1,\omega)$  with $\langle M_{\pm 1}\rangle=0$.  
{For simplicity, we restrict attention to leading-order terms as $\eta\rightarrow 0$, expanding (\ref{eq:short}) using (\ref{expansion}).}
{Expressions for $a_{\pm 1}$} give, in terms of the drop displacement $\mathcal{C}$ and width $\mathcal{W}$, 
{
\begin{align}
\label{eq:cdotwdot}
\dot{\mathcal{C}} =  -\tfrac{1}{2}{(M_{+1}-M_{-1}) \dot{a_0} }, \quad
\dot{\mathcal{W}} = 2 \dot{a_0}-{(M_{+1}+M_{-1}) \dot{a_0}}-{(6 \dot{a_0}/a_0)(\mathcal{W} -2 a_0)}.
\end{align}
}
{We set} $\mathcal{C}(t) = (M_{+1} - M_{-1}) \bar{\mathcal{C}}(t)$ and $\mathcal{W}(t) = 2 a_0 + (M_{+1} + M_{-1}) \bar{\mathcal{W}}(t)$, {where $\bar{\mathcal{C}}(0)=0$, $\bar{\mathcal{W}}(0)=0$.  Integrating (\ref{eq:cdotwdot})}, $\bar{\mathcal{C}}$ and $\bar{\mathcal{W}}$ satisfy the deterministic equations
{
\begin{align}
\label{eq:cbarwbar}
&{\bar{\mathcal{C}}} = \tfrac{1}{2}(1 -{a_0} ), \quad
\bar{\mathcal{W}}=\tfrac{1}{7}\left( a_0^{-6} - a_0\right).
\end{align}
}
Thus the mean drop centre and width satisfy
$\langle\mathcal{C}\rangle = \bar{\mathcal{C}}\left(\langle M_{+1} \rangle - \langle M_{-1} \rangle\right) = 0$, $\langle\mathcal{W}\rangle = 2a_0 + \bar{\mathcal{W}}(\langle M_{+1} \rangle + \langle M_{-1} \rangle ) = 2a_0$ {as expected, while the variances are}
\begin{subequations}\label{AsymMeanVar}
\begin{align}
\mathrm{Var}(\mathcal{C}) &= \bar{\mathcal{C}}^2\left(\mathrm{Var}(M_{+1}) + \mathrm{Var}(M_{-1}) - 2 \mathrm{Cov}(M_{+1}, M_{-1})\right)  
= 2\bar{\mathcal{C}}^2 \sigma^2 \left(1 - e^{-{2}/{l^2}} \right), \\
\mathrm{Var}(\mathcal{W}) &= \bar{\mathcal{W}}^2\left(\mathrm{Var}(M_{+1}) + \mathrm{Var}(M_{-1}) + 2 \mathrm{Cov}(M_{+1}, M_{-1})\right) 
 =2\bar{\mathcal{W}}^2 \sigma^2 \left(1 +  e^{-{2}/{l^2}}  \right),
\end{align}
\end{subequations}
using (\ref{eq:covar}) directly to evaluate covariances.
 
Predictions of the PDE system (\ref{PDE0}) , the ODE system (\ref{ODE}, \ref{eq:alpha}) and the explicit expressions (\ref{AsymMeanVar}) are compared in Figure~\ref{EOV}.  The linear dependence of $\sigma_{\mathcal{C}}$ and $\sigma_{\mathcal{W}}$ on $\sigma$ is reflected by simulations, except for larger variance where the assumption $M_{\pm}=\overline{M}(\pm1,0)$ breaks down because the effects of axial diffusion may also be significant.  The dependence on {the} correlation length is also captured well for $\ell \gtrsim 1$, but not at smaller $\ell$ where again the effects of axial diffusion are likely to become significant.  {The present predictions could be refined by incorporating $O(1/\ln\eta)$ corrections to $a_0$ in (\ref{loeq}), which would also include the influence of the initial viscosity distribution across the bulk of the drop through the integrals $\mathcal{I}_1$, $\mathcal{I}_2^\pm$ in (\ref{ODE}) and their correlations with $M_{\pm}$.  However we focus instead on incorporting the effects of axial diffusion.}

\section{An approximate model for axial diffusion}

Compression of the viscosity field at the contact line limits the time over which computations can be pursued (for fixed $Pe\gg1$), particularly when initial solute gradients are large.  Naturally, axial diffusion can be expected to have a growing influence in each compressive region as time increases.  We now develop a simple model to describe drop motion when diffusion is sufficient to homogenize the solute in the short compressive regions behind each contact line.  

We model the wedge behind the contact line, in which the flow is compressive, by taking $H\approx \theta_+(a_+-X)+\eta$ for $b_+ < X < a_+$ and $H=\eta$ for $X > a_+$, {where $b_+$ represents the rear boundary of the wedge, defined below}.  (For clarity we initially consider only the right-hand contact line.)  In this simple compartmental approach, $\theta_+$ represents the approximate contact angle at the edge of the outer region.  The inner region near the front of the wedge at $X=a_+$ has length $\eta/\theta_+$, which is short compared to $a_+$; we introduce $\varepsilon_+=\eta/(a_+\theta_+) \ll 1$.  Within the inner region $\overline{U}=\dot{a}_+(1-(\eta/H))$ (from (\ref{InVelocitya})); combining this with the stretching flow in the outer region gives a composite expression for the velocity across the wedge as
\begin{align}
\overline{U}=\dot{a}_+\left(1-\frac{\eta}{H}\right) +\dot{a}_+\left(\frac{X - a_-}{\mathcal{W}}-1\right) \qquad (b_+<X<a_+).
\end{align}
Thus $\overline{U}_X=0$ at $b_+=a_+ (\varepsilon_++1)-\sqrt{a_+ \varepsilon_+ \mathcal{W}}$, defining the rear of the wedge.  The fluid volume in the wedge, $V_+=\int_{b_+}^{a_+} H\,\mathrm{d}X$, satisfies, from (\ref{PDE0}a),
\begin{align}
\label{eq:vdot}
\dot{V}_+=(\dot{a}_+-\overline{U}\vert_{a_+})H\vert_{a_+} - (\dot{b}_+-\overline{U}\vert_{b_+})H\vert_{b_+}.
\end{align}
The mass of solute in the wedge, 
\begin{equation}
N_+=\int_{b_+}^{a_+} H\overline{M}\,\mathrm{d}X, 
\label{eq:solmass}
\end{equation}
satisfies (from (\ref{PDE0}b), neglecting diffusive fluxes at the edges of the wedge)
\begin{align}
\label{eq:ndot}
\dot{N}_+=(\dot{a}_+-\overline{U}\vert_{a_+})(H\overline{M})\vert_{a_+} - (\dot{b}_+-\overline{U}\vert_{b_+})(H\overline{M})\vert_{b_+}.
\end{align}
At the front of the wedge, $H=\eta$ and $\overline{U}=0$; at the rear, $H=\eta  \mathcal{W}/\sqrt{a_+ \varepsilon_+ \mathcal{W}}$ and $\overline{U}=\dot{a}_+ \left(\mathcal{W} + a_+ \varepsilon_+ - 2 \sqrt{a_+ \varepsilon_+ \mathcal{W}}\right)/\mathcal{W}$. We then assume that the compressed viscosity field is mixed by diffusion within the wedge, so that {the integral in (\ref{eq:solmass}) may be approximated as} $N_+\approx \widetilde{M}_+ V_+$ and $\overline{M} \vert_{b+}=\widetilde{M}_+$, in which case (\ref{eq:vdot}, \ref{eq:ndot}) give the leading-order {approximation}, as $\varepsilon_+ \to 0$, of the evolving solute concentration in the wedge as
\begin{align}
\label{eq:wedge}
\dot{\widetilde{M}}_{\pm}=\pm2\dot{a}_{\pm} (M^{\pm}-\widetilde{M}_{\pm})/\mathcal{W},
\end{align}
(treating the left-hand contact line analogously).  The reservoir of solute in the wedge is fed by a source in the film ahead, and diluted by expansion of the wedge.  Our candidate model for spreading, accounting for the effects of diffusion where the flow is strongly compressive, therefore uses (\ref{ODE}) with $\alpha_1=A_1$ and $M_{\pm}$ replaced by $\widetilde{M}_{\pm}(t)$, supplemented with (\ref{eq:wedge}).

The weak disorder limit of this model is particularly revealing.  Writing $\overline{M}=1+\overline{M}_1+\dots$, $a_{\pm}=\pm a_0+a_{\pm1}+\dots$ and $\widetilde{M}_{\pm} = 1 + \widetilde{M}_{\pm1} + \cdots$, (\ref{eq:wedge}) yields $\dot{\widetilde{M}}_{\pm1} = \dot{a}_0(\overline{M}_1(\pm a_0, 0) -\widetilde{M}_{\pm1} )/a_0$, $\mathcal{C} = (\widetilde{M}_{+1} - \widetilde{M}_{-1}) \bar{\mathcal{C}}$ and $\mathcal{W} = 2 a_0 + (\widetilde{M}_{+1} + \widetilde{M}_{-1}) \bar{\mathcal{W}}$, with $\bar{\mathcal{C}}$ and $\bar{\mathcal{W}}$ satisfying (\ref{eq:cbarwbar}) and $\overline{M}_1(X,0) = \mathcal{G}(X,\omega)$.  Taking $a_{0}(0)= 1$ and $\widetilde{M}_{\pm1}(0)=\overline{M}_1(\pm1, 0)$ leads to
\begin{align}
\widetilde{M}_{\pm1} (t) = \frac{1}{a_0}\left[\overline{M}_1(\pm1, 0) + \int_1^{a_0} \overline{M}_1(\pm X, 0)\,\mathrm{d}X\right].
\label{eq:mint}
\end{align}
This shows how the solute concentration in the wedge has contributions from the initial condition and from solute swept up by the advancing contact line.  As $a_0$ increases the latter component dominates the former, showing a fading memory of the initial condition.   Because $\widetilde{M}_1$ is a linear functional of the initial solute distribution, its statistics can be evaluated directly (Appendix~\ref{sec:wedge}) to give
\begin{subequations}\label{AsymMeanVarWedge}
\begin{align}
\langle\mathcal{C}\rangle =& 0, \\
\langle\mathcal{W}\rangle =&  2a_0,\\
\mathrm{Var}(\mathcal{C}) =& \frac{2\sigma^2\bar{\mathcal{C}}^2}{a_0^2}\Bigg[1 - 2 l^2 - \mathrm{e}^{-\frac{2}{l^2}} + l^2 \left(2 \mathrm{e}^{-\frac{(a_0-1)^2}{2 l^2}} - \mathrm{e}^{-\frac{2 a_0^2}{l^2}} + 2 \mathrm{e}^{-\frac{(a_0+1)^2}{2 l^2}} - \mathrm{e}^{-\frac{2}{l^2}}\right)\nonumber \\
&+ \sqrt{2 \pi } a_0 l \left(\mathrm{erf}\left(\frac{a_0-1}{\sqrt{2} l}\right) - \mathrm{erf}\left(\frac{\sqrt{2} a_0}{l}\right) + \mathrm{erf}\left(\frac{a_0+1}{\sqrt{2} l}\right)\right)\Bigg], \\
\mathrm{Var}(\mathcal{W}) =& \frac{2\sigma^2\bar{\mathcal{W}}^2}{a_0^2}\Bigg[1 - 2 l^2 + \mathrm{e}^{-\frac{2}{l^2}} + l^2 \left(2 \mathrm{e}^{-\frac{(a_0-1)^2}{2 l^2}} + \mathrm{e}^{-\frac{2 a_0^2}{l^2}} - 2 \mathrm{e}^{-\frac{(a_0+1)^2}{2 l^2}} + \mathrm{e}^{-\frac{2}{l^2}}\right)\nonumber \\
&+ \sqrt{2 \pi } a_0 l \left(\mathrm{erf}\left(\frac{a_0-1}{\sqrt{2} l}\right) + \mathrm{erf}\left(\frac{\sqrt{2} a_0}{l}\right) - \mathrm{erf}\left(\frac{a_0+1}{\sqrt{2} l}\right)\right)\Bigg].
\end{align}
\end{subequations}

{Figure~\ref{EOV} shows how the modified system captures the reduction in $\sigma_{\mathcal{C}}$ and $\sigma_{\mathcal{W}}$ as $\ell$ falls to zero.}   For $\ell=O(1)$, (\ref{AsymMeanVarWedge}) recovers (\ref{AsymMeanVar}) for $a_0\rightarrow 1$ while both variances are proportional to $\sqrt{2\pi}a_0\ell$ for $a_0 \gg 1$ at leading order.  Likewise both variances are proportional to $1+\sqrt{2\pi} a_0 \ell$ for $\ell \ll1$ with $a_0=O(1)$.   The increase of variance with $\ell$ when the correlation length is short compared to the drop radius can be explained as follows: diffusion acting within each wedge will tend to suppress the effect of fluctuations in the accumulated viscosity field; however increasing the correlation length suppresses this effect, in each contact line independently, promoting variation in drop location and width.  {The approximations (\ref{AsymMeanVar}) and (\ref{AsymMeanVarWedge}) indicate how variances change with time as the drop spreads (through their dependence on $a_0$), as long as the drop stays thick compared with the precursor film.}

\section{Discussion}

Complex liquids in natural environments can have spatially heterogeneous properties that influence, and are transported by, a flow.  Although diffusion can be expected to suppress spatial gradients over long timescales, heterogeneity will persist in liquids containing large molecular-weight structures with low mobility.  In practical applications, the heterogeneity can often be quantified at best at a statistical level, requiring flow outcomes to be described in terms of distributions.  The example we present here illustrates some of the challenges of this task.  Regions of strong compression quickly generate large spatial gradients in the transported material, far narrower than the physical boundary layers within the flow, which rapidly inflate computational cost; this cost is magnified by the requirement to simulate multiple realisations of the problem.  


The example we consider here, motivated by an application in respiratory physiology, illustrates the benefits (and limitations) of low-order approximations of the flow, which can be used to predict outcomes and their variability.  When heterogeneity is weak, drop spreading rates are determined primarily by conditions near each contact line at the start of the spreading process; the drop 'samples' restricted features of the initial viscosity distribution and these have long-lived influence.  In this case we were able to derive explicit expressions for the mean and variance of variables describing the drop's motion, in terms of parameters describing the structure of the initially heterogenous liquid.  A more complex picture emerges for a strongly heterogeneous liquid, for which spreading is inhibited by patches of elevated viscosity encountered by the advancing contact lines.   In this case we derived an \textit{ad hoc} model that shows how a reservoir of solute immediately behind the contact line regulates spreading rates over long timescales.

We have focused attention on a parameter range that is accessible to analysis.  A slender geometry allowed the use of lubrication theory {and the assumption of a fully wetting fluid interacting wtih a pre-wetted surface obviated the need to include disjoining pressure}.  We assumed a linear relationship between viscosity and the distribution of a passively transported solvent, and assumed that the solvent had a sufficiently large molecular weight for diffusion to suppress gradients across but not along the liquid layer.  We also assumed that the correlation length of the initial solute distribution exceeded the film thickness.  The resulting system of coupled nonlinear hyperbolic/parabolic PDEs generates solutions with large localised gradients requiring careful numerical treatment.   To gain physical insight we derived asymptotic approximations exploiting the difference between the height of the drop and the depth of the precursor film over which it spreads.  This yielded a {hierarchy of } algebraic/ODE systems for the location of the two contact lines, which can predict the mean and variance of drop width and lateral displacement.   Naturally many features arising in applications should be addressed in future studies, not least spreading in two spatial dimensions.

The ODE model revealed the mechanism whereby drop motion is arrested when encountering a patch of elevated viscosity.  The viscosity field is initially steepened within an inner layer near the contact line (of approximate width $\eta t^{2/7}$ at large times --- this is short compared to the drop width of approximate order $t^{1/7}$, assuming $t\ll \eta^{-7}$), leading to a ramp-like distribution passing slowly towards the rear of the inner region.  (The shear rate in the inner region is order $1/(t^{8/7}\eta)$, sufficiently to lead to exponentially rapid compression of the viscosity field.) As the more viscous liquid invades the inner region, the matching parameter $\alpha_{\pm}$ increases in magnitude (Figure~5), causing slowing of the contact line's motion.  Compression of the viscosity field extends into the overlap region between the inner and outer problems (a distance of order $(\eta t^{3/7})^{1/2}$ behind the contact line).   Thus, in the absence of longitudinal diffusion, characteristics are confined within this "compressive wedge."  In practice, diffusion can be expected to suppresses gradients in this wedge over large times.  In this case spreading rates are increasingly regulated by features of the viscosity field encountered during the spreading process.  The increasing influence of material encountered during spreading is neatly illustrated by (\ref{eq:mint}).

In terms of the application motivating this study, the primary insight concerns the role of the precursor (mucus) film in regulating the spreading of an inhaled drop of a different material.  Assuming the liquids are fully miscible, the drop will slowly accumulate endogenous material at its leading edge and fluctuations in its motion can therefore be associated directly with initial heterogeneities in the precursor film.  As Figure~\ref{EOV}(e) shows, fluctuations in drop location (relative to drop radius) are most pronounced when the drop radius is comparable to the correlation length of the viscosity distribution.  From a methodological perspective, our study shows how low-order physical models, combined with weak disorder expansions, can be effective in quantifying the statistical variability in flow outcomes.

\section*{Acknowledgements}

We are grateful to a referee for spotting an error in an earlier version of this work.  The authors have no competing interests.  Author contributions: OEJ and FX jointly conceived the study; FX performed numerical simulations; OEJ and FX jointly undertook the model development and analysis; OEJ and FX jointly drafted the manuscript.  Datasets from this study are available at http://dx.doi.org/10.5061/dryad.t2v1b.  This work was funded by EPSRC grant EP/K037145/1.

\begin{appendix}

\section*{Appendix}

\subsection*{(a) Model derivation}

We consider a liquid layer bounded below by a horizontal solid substrate and above by a free surface. We introduce Cartesian coordinates $(x ,z)$ such that the solid substrate lies on $z = 0$ and the free surface occupies $z = h(x, t)$, where $t$ is time. The liquid is incompressible, Newtonian and has dynamic viscosity $\mu(x, z, t)$, constant density $\rho$ and uniform surface tension $\sigma$.   The Reynolds number is assumed sufficiently small for inertia to be neglected, so that the flow field $(u, w)$ and pressure $p$ satisfy the Stokes equations, which when accounting for spatially varying viscosity may be expressed as
\begin{subequations}\label{emomentum2}
\begin{align}
\label{emass}
0&=u_x+ w_z, \\
\label{emomentum2_1}
0&= -p_x + \mu \left(u_{xx} + u_{zz}\right) + 2 \mu_x u_x+ \mu_z\left(u_z + w_x\right), \\
\label{emomentum2_2}
0&= -p_z + \mu \left(w_{xx} + w_{zz}\right) + 2 \mu_z w_z + \mu_x\left(u_z + w_x\right).
\end{align}
\end{subequations}
We assume the viscosity $\mu=\mu(c(x, z, t))$ is determined by the distribution of a chemical species with concentration $c(x, z, t)$, which satisfies the transport equation
\begin{align}\label{transport}
c_t + (c u)_x + (c w)_z = \gamma \left(c_{xx} + c_{zz}\right),
\end{align}
where  $\gamma$ is a constant diffusivity.  No-flux conditions are imposed on $c$ at $z=0$ and $z=h$. 

We introduce a length scale $l_0$, height scale $h_0$ (defining $\epsilon \equiv {h_0}/{l_0} \ll 1$), viscosity scale $\mu_0$, velocity scale $u_0=\epsilon^3 \sigma/\mu_0$ and concentration scale $c_0$, and nondimensionalize variables using $(x, z, h) = h_0 ({X}/{\epsilon}, Z, H )$, $(u,w) = u_0 (U, \epsilon W)$, $t = ({h_0}/{\epsilon u_0}) T$,  $p = ({\mu_0 u_0}/{\epsilon h_0}) P$, with $\mu = \mu_0 M(C)$ and $c = c_0 C$, where the function $M(C)$ is to be specified.  We define the P\'{e}clet number $Pe = {l_0 u_0}/{\gamma}$.

 The governing equations and boundary conditions become
\begin{subequations}\label{dimensionless3}
\begin{align}
\label{dimensionless3_1}
&U_X+W_Z= 0, \quad 
0 = -P_X + (M U_Z)_Z , \quad
0= - P_Z , \\
\label{dimensionless3_4}
&C_T + (C U)_X + (C W)_Z = \frac{1}{Pe}C_{XX} + \frac{1}{\epsilon^2 Pe}C_{ZZ}, \\
\label{dimensionless3_5}
&W = 0, \quad U = 0, \quad C_Z = 0, \quad (Z = 0), \\
\label{dimensionless3_7}
&W = H_T + U H_X, \quad P  = -H_{XX} , \quad U_Z=  0 , \quad C_Z = \epsilon^2 H_X C_X,\quad (Z = H(X, T)),
\end{align}
\end{subequations}
with $M=M(C)$.   Here we have eliminated terms of $O(\epsilon^2)$ from the flow equations (as is standard in lubrication theory) but have retained all terms in the solute transport equation.  It follows that $P(X,T) = - H_{XX}$, and integration of the horizontal momentum equation yields
\begin{align}\label{integration_momentum}
U(X, Z, T) =  P_X\left( \int_0^Z\frac{Z'dZ'}{M(X,Z',T)} - \int_0^Z\frac{HdZ'}{M(X,Z',T)} \right).
\end{align}

We define an averaging operator on the function $\Phi(X, Z, T)$  as
\begin{align}\label{average_operator}
&\overline{\Phi} \equiv \frac{1}{H}\int_0^H \Phi(X, Z, T) dZ,
\end{align}
so that $\overline{\Phi_Z} = \left(\Phi|_{Z=H} - \Phi|_{Z = 0}\right)/H$ and
\begin{align}
&\overline{ \Phi_X} = {\overline{\Phi}}_X + \frac{1}{H}H_X\left(\overline{\Phi} -\Phi|_{Z=H}\right),  
\quad \overline{\Phi_T} = {\overline{\Phi}}_T + \frac{1}{H}H_T\left(\overline{\Phi} -\Phi|_{Z=H}\right).
\end{align}
Exploiting these identities, averaging the mass conservation equation (\ref{dimensionless3_1}-1) and using the boundary conditions (\ref{dimensionless3_5}-1) and (\ref{dimensionless3_7}-1) yields $H_T + \left(H\overline{ U}\right)_X   = 0$, as in (\ref{PDE0}a). Likewise, averaging the transport equation (\ref{dimensionless3_4}) and imposing boundary conditions (\ref{dimensionless3_5}-1, \ref{dimensionless3_7}-1, \ref{dimensionless3_5}-3, \ref{dimensionless3_7}-4) gives 
\begin{align}\label{transport3}
\left(H\overline{C}\right)_T + \left(H\overline{ C U}\right)_ X = {Pe}^{-1}\left(H\overline{ C_X}\right)_X.
\end{align}
We may combine {(\ref{PDE0}a, \ref{transport3})} to give 
\begin{align}\label{transport4}
{\overline{ C}}_T  + (\overline{C U})_ X - \overline{ C}\,{\overline{ U}}_ X + \frac{1}{H}H_X \left(\overline{ CU} -\overline{ C}\,\overline{ U} \right) = \frac{1}{Pe}\frac{(H\overline{ C_X})_X}{H}.
\end{align}

We introduce the decomposition
\begin{align}\label{decomp}
&C = \overline{C} + C', \quad U = \overline{U} + U', \quad W = \overline{W} + W', \quad M = \overline{M} + M',
\end{align}
with $\overline{C}=\overline{C}(X,T)$ and $\overline{C'}\equiv 0$ etc. Averaging (\ref{dimensionless3_4}) and using the decomposition (\ref{decomp}) gives
\begin{align}
\label{dte}
 {\overline{ C}}_T  + C'_T +  (\overline{C}\, \overline{U})_X &+ \overline{C} U'_X + U' {\overline{C}}_X + (\overline{U} C')_X + (C' U')_X\nonumber \\
&+ \overline{C}  W'_Z + \overline{W} C'_Z + (C' W')_Z = \frac{1}{Pe}{\overline{C}}_{ XX} + \frac{1}{Pe}C'_{XX} + \frac{1}{\epsilon^2 Pe}C'_{ZZ}, 
\end{align}
while the cross-sectional averaged transport equation (\ref{transport4}) becomes
\begin{align}\label{date}
&{\overline{C}}_T   + \overline{U}\, \overline{C}_ X+  (\overline{C' U'})_X  + \frac{1}{H}H_X \overline{C' U'} 
= \frac{1}{Pe}{\overline{C}}_{ XX} + \frac{1}{Pe }\frac{H_X \overline{C}_X}{ H} + \frac{1}{Pe}\frac{(H\overline{C'_X})_X}{H}.
\end{align}
Subtracting (\ref{date}) from (\ref{dte}) gives
\begin{align}\label{feq} 
& C'_T  + U' {\overline{C}}_X + (\overline{U} C')_X + (C' U')_X -  \left(\overline{C' U'}\right)_X - \frac{1}{H}H_X \overline{ C' U'} + \overline{W} C'_Z +  (C' W')_Z \nonumber \\
&= \frac{1}{Pe}C'_{XX} + \frac{1}{\epsilon^2 Pe}C'_{ZZ} - \frac{1}{Pe }\frac{H_X {\overline{C}}_X}{ H} - \frac{1}{Pe}\frac{(H\overline{ (C'_X)})_X}{H}.
\end{align}

We seek the limit in which $C' \ll \overline{ C} \sim 1$ while $U' \sim \overline{ U }\sim 1$, taking $Pe \gg 1$ as $\epsilon \to 0$.  Anticipating the dominant balance in (\ref{feq}) to be $U' {\overline{C}}_X = C'_{ZZ}/{\epsilon^2 Pe}$, so that $C' =O( \epsilon^2 Pe)$, the terms in (\ref{date}) fall into four categories with magnitude $O(1)$ (advection), $O(1/Pe)$ (diffusion), $O(\epsilon^2 Pe)$ (Taylor dispersion) and $O(\epsilon^2)$.  Thus for  $1/{\epsilon} \gg Pe \gg 1$, the approximation of (\ref{date}) up to  $O(1/Pe)$ is
\begin{align}\label{transport5}
&{\overline{ C}}_T  + \overline{U}\,{\overline{ C}}_X    = \frac{1}{Pe}\frac{(H{\overline{ C}}_X)_X}{H}.
\end{align}
We retain the $O(1/Pe)$ contribution of diffusion in (\ref{transport5}) to facilitate numerical simulations.  Assuming that the viscosity $M$ linearly depends on $C$ yields (\ref{PDE0}b) and $M' \ll \overline{ M}$.  Then, averaging the horizontal velocity component  (\ref{integration_momentum}) gives
\begin{align}\label{ave_h_vel}
\overline{ U} =  &
P_X\left( \frac{1}{H}\int_0^H\int_0^Z\frac{Z'dZ'dZ}{M(X,Z',T)} - \int_0^H\int_0^Z\frac{dZ'dZ}{M(X,Z',T)} \right) \approx  - P_X \frac{H^2}{3\overline{ M}}  .
\end{align}
with error $O(1/Pe, \epsilon^2 Pe, \epsilon^2)$, as in (\ref{PDE0}a).

{A similar formulation has been adopted by Karapetsas \hbox{et al.} \cite{karapetsas2016} in a study of thin-film suspension flow, for which a nonlinear relation between viscosity and particle concentration was retained.  As in that study, we assume here that the solute field does not influence the interfacial tension; for suspensions, linearisation of $M(C)$ is appropriate at low volume fractions \cite{stickel2005}.}
 
\subsection*{(b) {Karhunen--Lo\'{e}ve decomposition}}

We use the Karhunen--Lo\'{e}ve decomposition to sample the Gaussian random field $\mathcal{G}$, which then gives one sample of $\mathcal{M}$ via exponentiation. Given the spatial grid $\{X_i\}$, $i = 0,1, \cdots, n$, on the computational domain $[-L, L]$, the covariance function $k_\mathcal{G}$ produces a covariance matrix $K= \{k_\mathcal{G}(X_i, X_j)\}$, $i, j = 0, 1, \cdots, n$, which can be factorised as $K = V \Lambda V^T$, where $\Lambda$ is the $(n+1) \times (n+1)$ diagonal matrix of eigenvalues , $\lambda_0 \geq \lambda_1 \geq \cdots \geq \lambda_n \geq 0$, of $K$, and $V = [\bm{v}_0, \cdots, \bm{v}_n]$ is the $(n+1) \times (n+1)$ matrix  whose columns $\bm{v}_i$ are the eigenvectors of $K$.  To resolve the $M$ field properly, we choose the grid width such that it is smaller than one fifth of the correlation length $l$. As in \cite{Lord2014}, the discrete random field $\bm{\mathcal{G}} := [\mathcal{G}(X_0,\omega), \cdots, \mathcal{G}(X_n,\omega)]^T$  can be generated as
\begin{align}\label{KLDecom}
\bm{\mathcal{G}} = \sum_{i = 0}^n \sqrt{\lambda_i} \bm{v}_i \xi_i,
\end{align}
where $\xi_i$ are independent and identically distributed Gaussian random variables with zero mean and unit variance. For large $n$, we further truncate the sum in (\ref{KLDecom}) after $n'$ ($\ll n$) terms and use the approximate discrete random field
\begin{align}\label{TrunKLDecom}
&\bm{\mathcal{G}}' = \sum_{i = 0}^{n'} \sqrt{\lambda_i} \bm{v}_i \xi_i.
\end{align}
The quality of the approximation of $\bm{\mathcal{G}}' \approx \bm{\mathcal{G}}$ is determined by the sizes of the neglected eigenvalues $\lambda_{n' + 1}, \cdots, \lambda_n$. Here we choose smallest $n'$ such that $\lambda_{n'}/\lambda_0 <10^{-3}$. 
  
\subsection*{(c) Compressive wedge approximation in the weak disorder limit}
\label{sec:wedge}

Writing $Y(X)\equiv \overline{M}_1(X,0)\pm\overline{M}_1(-X,0)$, (\ref{eq:mint}) implies
\begin{align}
\label{integral_of_grf}
\widetilde{M}_{+1}\pm\widetilde{M}_{-1}=\frac{1}{a_0} \left[Y(1)
+ \int_1^{a_0} Y(X)
\,\mathrm{d}X\right] \approx  \frac{1}{a_0} \left[Y(X_1) + \Delta X \sum_{j =1}^N Y(X_j)\right],
\end{align}
approximating the integral as a Riemann sum with $\Delta X = (a_0 - 1)/N$, $X_j = 1+(j-1)\Delta X $ and $N$ a large positive integer.
Clearly $\langle\widetilde{M}_{+1}\pm\widetilde{M}_{-1}\rangle = 0$ while
\begin{multline}
\mathrm{Var}(\widetilde{M}_{+1}\pm\widetilde{M}_{-1}) \approx\frac{1}{a_0^2} \left[\mathrm{Var}(Y(X_1)) + 2 \Delta X \sum_{j =1}^N \mathrm{Cov}(Y(X_1), Y(X_j)) \right. \\ \left. + (\Delta X)^2 \sum_{j =1}^N \sum_{i =1}^N \mathrm{Cov}(Y(X_i), Y(X_j))\right],
\end{multline}
and
\begin{align}
\mathrm{Cov}(Y(X_i), Y(X_j)) = & \mathrm{Cov}(\overline{M}_1(X_i,0)\pm \overline{M}_1(-X_i,0), \overline{M}_1(X_j,0)\pm \overline{M}_1(-X_j,0)) \nonumber \\
 = & \mathrm{Cov}(\overline{M}_1(X_i,0), \overline{M}_1(X_j,0)) \pm \mathrm{Cov}(\overline{M}_1(X_i,0), \overline{M}_1(-X_j,0)) \nonumber \\
&\pm \mathrm{Cov}(\overline{M}_1(-X_i,0), \overline{M}_1(X_j,0)) + \mathrm{Cov}(\overline{M}_1(-X_i,0), \overline{M}_1(-X_j,0)) \nonumber \\
=& 2\sigma^2\exp\left(-\frac{1}{2}\left(\frac{X_i-X_j}{l}\right)^2\right) \pm 2\sigma^2\exp\left(-\frac{1}{2}\left(\frac{X_i+X_j}{l}\right)^2\right).
\end{align}
Restoring sums to integrals leads to
\begin{align}
\label{integral_in_variance}
\mathrm{Var}(\widetilde{M}_{+1}\pm\widetilde{M}_{-1})  = & \frac{2\sigma^2}{a_0^2} \Bigg[ 1\pm \exp\left(-\frac{2}{l^2}\right) + 2\int_1^{a_0} \exp\left(-\frac{1}{2}\left(\frac{1-X}{l}\right)^2\right) \pm \exp\left(-\frac{1}{2}\left(\frac{1+X}{l}\right)^2\right)\,\mathrm{d}X \nonumber \\
& +\int_1^{a_0} \int_1^{a_0} \exp\left(-\frac{1}{2}\left(\frac{X-X'}{l}\right)^2\right) \pm \exp\left(-\frac{1}{2}\left(\frac{X+X'}{l}\right)^2\right)\,\mathrm{d}X \,\mathrm{d}X'\Bigg],
\end{align}
with Riemann sums converted to integrals as $N \to \infty$.  The integrals in (\ref{integral_in_variance}) can be explicitly evaluated 
in terms of the error function to give (\ref{AsymMeanVarWedge}).

\end{appendix}

\end{document}